\documentclass[twocolumn,amsmath,amssymb,showpacs,superscriptaddress,prl]{revtex4}

\usepackage{braket}
\usepackage{graphicx}
\usepackage{epstopdf}
\usepackage{appendix}
\usepackage{graphicx}
\usepackage{color}

\begin{document}

\title{Topological classification of ${\bf k \cdot p}$ Hamiltonians for Chern insulators}

\author{Frank Kirtschig}
\affiliation{Institute for Theoretical Solid State Physics, IFW Dresden, PF 270116, 01171 Dresden, Germany}
\author{Jeroen van den Brink}
\affiliation{Institute for Theoretical Solid State Physics, IFW Dresden, PF 270116, 01171 Dresden, Germany}
\affiliation{Department of Physics, Dresden University of Technology, 01062 Dresden, Germany}
\author{Carmine Ortix}
\affiliation{Institute for Theoretical Solid State Physics, IFW Dresden, PF 270116, 01171 Dresden, Germany}

\date{\today}

\begin{abstract}
We proof the existence of two different topological classes of low-energy ${\bf k \cdot p}$ Hamiltonians for Chern insulators. 
Using the paradigmatic example of single-valley two-band models,
we show that  ${\bf k \cdot p}$ Hamiltonians that we dub {\it local} have a topological invariant corresponding precisely to the Hall conductivity and linearly dispersing chiral midgap edge states at the expansion point.  {\it Non-local}  ${\bf k \cdot p}$ Hamiltonians have a topological invariant that is twice the Hall conductivity of the system. 
This class is characterized by a non-local bulk-edge correspondence with midgap edge states appearing
away from the high-symmetry ${\bf k \cdot p}$ expansion point.
\end{abstract}

\pacs{73.43.-f, 73.20.-r, 03.65.Vf}

\maketitle

\paragraph{Introduction -- } 
The discovery of topological insulators (TIs)  \cite{has10,moo10}  has brought to light a new state of quantum matter that has had a tremendous impact in the field of fundamental condensed matter physics as well as for potential application in spintronics and quantum computation. The TIs are insulating in the bulk, but they do possess metallic edge states \cite{has10,moo10,hal82,kane05,wu06}, which are topologically protected against generic perturbations preserving the symmetries of the underlying topological class \cite{alt06} and the intrinsic insulating behavior. 
Most importantly, the robustness of these edge states is encoded in a topological invariant classifying the ground state of the system: this is the celebrated bulk-edge correspondence \cite{lau81,ess11}. 
With non-interacting lattice Hamiltonians at work, the band structure of an insulator can be viewed as a mapping from the periodic Brillouin zone, which has the topology of a torus $\mathcal{T}^d$ in $d$ dimensions, to the space of Bloch Hamiltonians with an energy gap. In two-dimensional (2D) systems, this mapping, in turn,  allows for the definition of a topological invariant, which is referred to as the Chern number $C$, without invoking additional symmetries such as time-reversal \cite{kane05} or point-group symmetries of the lattice \cite{fu11}.  

This  becomes manifest in a simple fashion by particularizing to the situation of a two-band Hamiltonian, which can be represented in the space of Pauli matrices $\boldsymbol{\tau}$ as 
${\cal H}\left(\mathbf{k}\right)=\varepsilon(\mathbf{k}) {\cal I}_2 - \mathbf{d}(\mathbf{k})\cdot\boldsymbol{\tau}$,
where, since $\varepsilon(\mathbf{k})$ multiplies the identity matrix ${\cal I}_2$, the eigenstate structure, and consequently the topological character of the electronic ground state of the system, depends only on $\mathbf{d}=(d_x,d_y,d_z)$.
The Chern number of the Hamiltonian above is given by 
\begin{equation}
C= \dfrac{1}{4\pi}\int dk_x dk_y \, \hat{\mathbf{d}}\cdot(\partial_{k_x}\hat{\mathbf{d}}\times\partial_{k_y}\hat{\mathbf{d}})
\label{eq:chern}
\end{equation}
where $\hat{\mathbf{d}}$ is the normalized $\mathbf{d}$ vector of unit length, which lives on a sphere and thus defines a mapping from the 2-torus $\mathcal{T}^2$ of the Brillouin zone to the 2-sphere $\mathcal{S}^2$. With this, $C$ counts the number of times the image of the mapping wraps around the sphere, which is obviously an integer, cannot change 
under smooth deformations of the mapping, and thus defines a topological invariant of the system. 

Instead of the full electronic band structure over the entire BZ one often uses the effective, long-wavelength description of a system: 
its ${\bf k \cdot p}$  Hamiltonian \cite{win03} that corresponds to the continuum description of its low-energy bands. The obvious question that arises is whether and how the topological invariant classifying a Chern insulator can be defined in such an effective continuum description of the system. 
This is indeed particularly timely in view of the intense research effort on closely-related topological states of matter in inverted semiconductors, such as HgTe quantum wells \cite{bhz06}, or bismuth-based materials \cite{zha09}. 
On top of this, effective ${\bf k \cdot p}$ theories have been shown to be essential in order to gain insights into the fundamental electronic properties of the topologically protected surface or edge states \cite{fu09,ort12,ca13}. In this Letter, we address this question and provide a classification of low-energy, long-wavelength Hamiltonian for 2D Chern insulators, thereby identifying two different topological classes defined in terms of topological invariants different in nature. 
The {\it local}  ${\bf k \cdot p}$ topological class  corresponds to the known case of low-energy continuum theories with the momentum space one-point compactified 
to a 2-sphere ${\mathcal S}^2$. The topological information is then contained in a ${\mathcal S}^2 \rightarrow {\mathcal S}^2$ mapping -- with the topological invariant
equal to the Chern number -- and manifested in the ubiquitous appearance of midgap linearly dispersing chiral edge states  with left-movers and right-movers crossing each other precisely at the ${\bf k \cdot p}$ expansion point. 
In the second  {\it non-local} ${\bf k \cdot p}$ class, the momentum manifold is instead holomorphic to a {\it non-compact} hemisphere. 
While this feature has  previously been associated with marginal topological behavior \cite{volo03} and a consequent absence of topological well-defined quantities  \cite{li10,li12}, we show that
a {\it one-dimensional} topologically non-trivial mapping from a circle to a circle is still allowed. The ensuing topological invariant then corresponds to the topological charge of the equator vortex and amounts to twice the Hall conductivity of the system. An explicit solution of the low-energy theory in semi-infinite systems shows,  {\it independent} of the boundary conditions, the appearance of 
edge modes at the top (bottom) of the valence (conduction) band and vanishing Fermi velocity at the ${\bf k \cdot p}$ expansion point.
The non-trivial topology then yields a {\it non-local} bulk-edge correspondence:  although the bulk topological invariant can be read off from the properties of low-energy electrons,  chiral edge states of finite Fermi velocity connecting the valence to the conduction band are encountered at BZ points away from  the ${\bf k \cdot p}$ expansion point.

\begin{figure}
\includegraphics[width=\columnwidth]{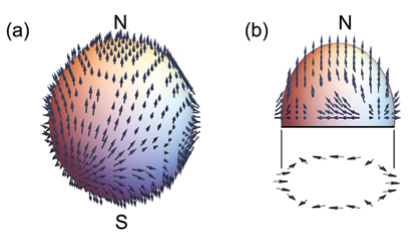}
\caption{(Color online) Top view patterns of the  $\hat{d}$ vector configurations on a Bloch sphere( {\it local} ${\bf k \cdot p}$ topological class (a)) and a Bloch hemisphere ({\it non-local} ${\bf k \cdot p}$ topological class (b)). The two topological classes are defined in terms of different topological invariants.}
\label{fig:map}
\end{figure}

\paragraph{ Local  ${\bf k \cdot p}$  topological class -- } 
When dealing with a 2D long-wavelength Hamiltonian,  the momentum manifold defining the band structure of an insulator corresponds to the real plane $\mathcal{R}^2$, which is known to be an element of the  trivial class \cite{cho82} -- mappings between the non-compact $\mathcal{R}^2$ and the compact 2-sphere $\mathcal{S}^2$ are generally topologically trivial \cite{li10} .  
A loophole for this statement appears whenever the infinite ${\bf k}$ plane can be compactified as a Riemann sphere \cite{li12}. Put in simple terms, this occurs if the normalized $\mathbf{d}$ vector converges in the ${\bf k} \rightarrow \infty$ limit to a single point of $\mathcal{S}^2$, say the south pole, regardless of the direction along which infinity is approached, {\it i.e.} $\lim_{|\mathbf{k}|\to\infty}\hat{\mathbf{d}}=\mbox{const}$. Mappings from this one-point compactified ${\bf k}$ plane to the 2-sphere subtended by $\hat{\mathbf{d}}$ are topologically well-defined, thereby allowing for a safe definition of the Chern number. To show this, we consider the example of the Bernevig-Hughes-Zhang (BHZ) model \cite{bhz06} originally introduced to discuss the time-reversal Quantum Spin Hall insulating phase in HgTe/CdTe quantum wells. Assuming a $U(1)$ spin symmetry, each of the two spin sectors of the BHZ Hamiltonian corresponds to a long-wavelength ${\bf k \cdot p}$ theory of a Chern insulator with the ${\mathbf{d}}$ vector reading
\begin{equation}
\mathbf{d}=(s_z k_x, k_y, m- k^2),
\label{eq:BHZkp}
\end{equation}
where  $s_z=\pm 1 $  distinguishes the two spin sectors. 
For $m>0$ the normalized $\mathbf{d}$ vector visits both the south and the north pole, wraps the unit sphere once, thereby implying an integer Chern number $C=s_z$. 
In the opposite $m<0$ regime instead, the image of the mapping does not cover the full sphere yielding a trivial $C=0$ Chern number. 
The well-defined topological properties of this long-wavelength Hamiltonian are reflected in the ubiquitous emergence of topologically protected edge states within the ${\bf k \cdot p}$
framework. 
To show this, we consider the long-wavelength model Hamiltonian defined on the half-plane given by $x>0$ and consider boundary conditions that ensure a self-adjoint extension of the half-plane Hamiltonian, {\it i.e.} $\braket{\Psi |{\cal H} \Phi} - \braket{{\cal H}^{\dagger} \Psi | \Phi} \equiv 0$ with the domains of ${\cal H}$ and ${\cal H}^\dagger$  equal. Those correspond to the usual Dirichlet fixed boundary conditions (FBC) \cite{liu10}, which clamp the wave function to zero at the boundaries, and to  "natural" boundary conditions (NBC), that can be also derived by minimizing a variational energy functional \cite{med12}, explicitly reading  $\partial{\cal H} / \partial {k_x} (k_x\to-i\partial_x) \psi|_{x=0} \equiv 0$. 

\begin{figure}
\includegraphics[width=\columnwidth]{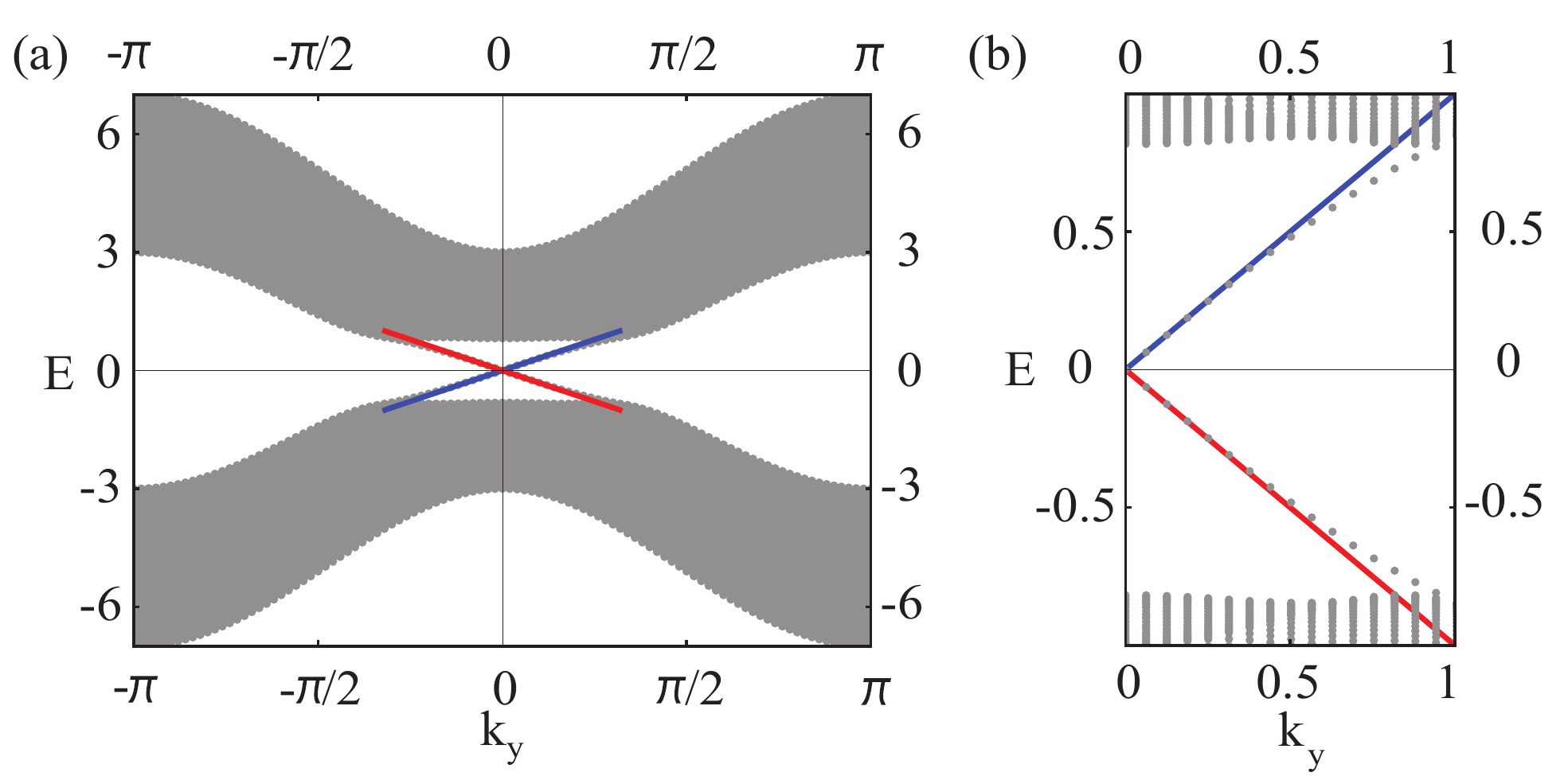}
\caption{(Color online) (a) Energy bands for the BHZ tight-binding model of Ref.~\onlinecite{bhz06}  in a ribbon of width $W=200 a$ with  $a$ the square lattice constant. The thick lines represent the analytical results for the dispersion of the topologically protected edge states as obtained from the low-energy ${\bf k \cdot p}$ expansion at the centre of the BZ. (b) Enlargement in the bulk gap region showing a perfect agreement between the analytical and the numerical results.} 
\label{fig:BHZ}
\end{figure}

We then look for topologically protected edge states  by assuming the ansatz for the localized wavefunction $\psi=e^{ik_y y}\sum_{i}\chi_{\lambda_i} e^{\lambda_i x}$ with $\mathfrak{R}(\lambda_i)<0$. By solving the corresponding Schr\"odinger equation, we obtain midgap chiral edge states of opposite chirality in the two spin sectors with energy $E=- s_z k_y$.  The corresponding wavefunctions read 
\begin{equation}
\Psi\propto\left [ i \, , -s_z \right]^T (e^{\lambda{_1} \, x}\pm e^{\lambda_{2} \, x})
\label{eq:edgestateBHZ}
\end{equation}
where the $\pm$ are for the NBC and the FBC boundary conditions respectively and the $\lambda_{1,2}$ satisfy
\begin{equation}
\lambda_{1,2}=-\frac{1}{2}\pm\sqrt{\frac{1}{4 }+k_y^2- m }, 
\label{eq:edgestateBHZlambda}
\end{equation}
which imply a renormalizability of the edge states wavefunction in the bulk gap region $m > k_y^2$. 
The dispersion of the edge states is shown with the thick lines in Fig.~\ref{fig:BHZ}. We find an excellent agreement with the numerical results obtained by solving the square lattice tight-binding Hamiltonian of Ref.~\onlinecite{bhz06} in a ribbon geometry, explicitly proving that the absence of a short-distance cutoff in the long-wavelength Hamiltonian {\it does not} change neither the topological properties nor the electronic characteristics of the topologically protected edge states.

\paragraph{Non-local  ${\bf k \cdot p}$ topological class --}
As a matter of fact, for certain Chern insulating lattice models a one-point compactification of the effective low-energy ${\bf k \cdot p}$  theory can strongly modify the intrinsic topological properties. This can be explicitly shown by considering a lattice model for spinless fermions on the checkerboard lattice \cite{kiv09} in the quantum anomalous Hall (QAH)
state \cite{hal88}. The corresponding $\mathbf{d}$ vector of the lattice Hamiltonian can be written as \cite{kou14}
\begin{equation}
\mathbf{d}=\left(\begin{array}{c} 4t \cos{\phi} \cos{\frac{k_x}{2}} \cos{\frac{k_y}{2}} \\ 4t \sin\phi \sin{\frac{k_x}{2}} \sin{\frac{k_y}{2}} \\ 2t_2 (\cos{k_x}-\cos{k_y})\end{array}\right)
\label{eq:checkerboard}
\end{equation}
where $t$ and $t_2$ are the nearest-neighbor and next-nearest-neighbor hopping amplitudes, while $\pm \phi$ are local magnetic fluxes picked up by the electrons while hopping around each plaquette  clock and aniclockwise respectively. Without loss of generality, we will assume in the following $t,t_2 >0$ and $0<\phi<\pi/2$. 
The effective low-energy, long-wavelength Hamiltonian can be derived expanding Eq.~\ref{eq:checkerboard} close to the corner of the Brillouin zone $M= \left\{\pi, \pi \right\}$  as 
\begin{equation}
\mathbf{d} \simeq (t\cos\phi k_x k_y,\Delta({\bf k}) ,t_2 (k_x^2-k_y^2)),
\label{eq:checkerboardkp}
\end{equation} 
where the time-reversal symmetry breaking mass $\Delta({\bf k})=\Delta_0 - \Delta_1{\bf k}^2$. 
Due to the absence of a compact ${\bf k}$-space, the mapping $\hat{\mathbf{d}}$ is topologically trivial, reflected in the fact that the ensuing Chern number takes a  meaningless non-integer value as long as $\Delta_1 \neq 0$.

\begin{figure}
\includegraphics[width=0.7\columnwidth]{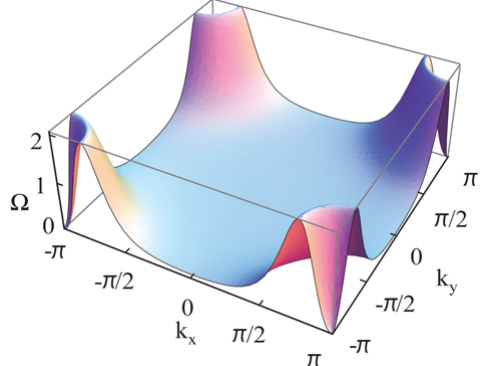}
\caption{(Color online) Plot of the integrand of Eq.~\ref{eq:chern} as a function of momentum ${\bf k}$ over the entire Brillouin zone, obtained from the checkerboard lattice model in the quantum anomalous Hall state with the parameter $t_2/t=1/2$ and $\phi=1/10$. The integrand peaks close to the $M$ points.} 
\label{fig:Berry}
\end{figure}

We therefore first seek for a natural one-point compactification by assuming a modified functional form of the time-reversal symmetry breaking mass $\Delta({\bf k})=\Delta_0 - \Delta_1{\bf k}^2 - \Delta_2{\bf k}^4$. Since in this case $\hat{\mathbf{d}}$ defines a topological non-trivial mapping, integers Chern numbers have to be restored. And indeed we find $C \equiv \Delta_0 / |\Delta_0| + \Delta_2 / |\Delta_2| \in\{\pm 2,0\}$  as easily follows from the fact that the $d$-wave symmetry of the effective low-energy Hamiltonian implies that the unit sphere subtended by $\hat{\mathbf{d}}$ can only be wrapped twice. 
This result is in striking contrast with the analysis of the checkerboard lattice model Eq.~\ref{eq:checkerboard} , which predicts $C=\pm 1$, even though the integrand in Eq.~\ref{eq:chern} is strongly localized close to the $M$ point of the Brillouin zone [c.f. Fig.\ref{fig:Berry}]. 

Having established that the topological characteristics of the checkerboard lattice model cannot be correctly described within a one-point compactified ${\bf k \cdot p}$ Hamiltonian, we now proceed to introduce the second topological class of effective long-wavelength Hamiltonians for Chern insulators and the topological invariant that can be defined thereof. 
We start out by requiring the momentum manifold $\in \mathcal{R}^2$ to be holomorphic to a non-compact hemisphere. 
This can be accomplished if in the ${\bf k} \rightarrow \infty$ limit the normalized $\mathbf{d}$ vector is planar and thus confined to the equator of a 2-sphere $\mathcal{S}^2$. 
It then follows that for a closed loop in momentum space in the large $|{\bf k}|$ limit, the planar $\mathbf{d}$ vector  defines a topologically non-trivial mapping from a circle $\mathcal{S}^1$ to a circle $\mathcal{S}^1$, whose ensuing  topological invariant counts the number of times $\hat{\mathbf{d}}$ encircles the origin. 

In order to show how this winding number is related to a well defined physical response,
we recall that the Hall conductivity $\sigma_H$ in units of $e^2/ h $ can be written as a line integral over a closed loop ${\cal C}$ in  momentum space
\begin{equation}
\sigma_H = \frac{1}{2\pi}\oint_{\mathcal{C}} d{\bf k}\cdot{\mathcal A} ({\bf k}),
\label{eq:hallcheckerboard}
\end{equation}
where the Berry connection of the occupied lower band ${\mathcal A} ({\bf k})= -i\bra{u({\bf k})} {\mathbf \nabla} \ket{u({\bf k})}$ in terms of the Bloch wavefunctions $\ket{u({\bf k})}$ \cite{hal04}. Since the Berry phase in Eq.~\ref{eq:hallcheckerboard} is gauge dependent, however, the gauge needs to be fixed. We accomplish this by requiring that the wavefunctions at the ${\hat d}_z \equiv 0$ equator are able to contract into the north or south pole. With this, it follows that  the lower band eigenstate at the ${\hat d}_z \equiv 0$ equator has to be written as $u({\bf k})=(1,{\hat d}_x + i {\hat d}_y)^T /\sqrt{2}$ for ${\hat d}_3>0$ and $u({\bf k})=({\hat d}_x - i {\hat d}_y, 1)^T /\sqrt{2}$ for ${\hat d}_3<0$. 
This gauge fixing procedure, in turns, yields $\sigma_H = W \hat{d}_z / (2 |\hat{d}_z|)$, where $W= 1/(2\pi)\oint_{\mathcal{C}}d {\bf k}\cdot[\hat{d}_x {\mathbf \nabla} \hat{d}_y - {\hat d}_y {\mathbf \nabla} {\hat d}_x ]  $ is a topological invariant corresponding to the topological charge of the $\hat{\mathbf{d}}$ vortex \cite{sun06}.

For the effective low-energy theory of the checkerboard lattice model Eq.~\ref{eq:checkerboardkp}, the momentum manifold can be rendered to be holomorphic to a hemisphere by neglecting the momentum dependence  of the time-reversal symmetry breaking mass $\Delta({\bf k}) = \Delta_0$. With this, it follows that the Hall conductivity $\sigma_H =  \Delta_0 / |\Delta_0|$, which is in perfect agreement with analysis of the lattice Hamiltonian, and reconciles the topological properties of the long-wavelength theory with the short-distance cutoff one. 

\begin{figure}
\includegraphics[width=\columnwidth]{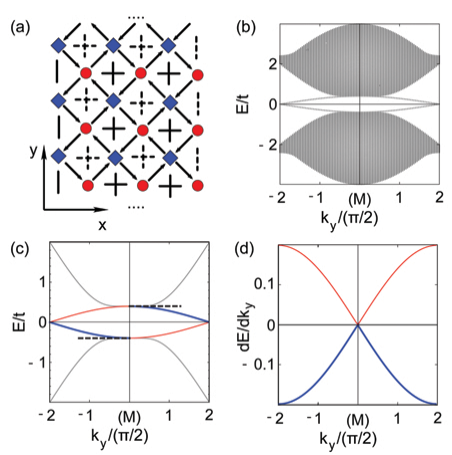}
\caption{(Color online) (a) A checkerboard lattice  (b) Energy bands for  ribbons in the QAH state with $t_2/t=1.2$, $\phi=1/10$. The edge state crossing always occurs away from the low-energy  ${\bf k \cdot p}$ expansion point M=$(\pi,\pi)$ where the bulk band gap is smallest. (c) Enlargement close to the $M$ point projection. The dashed line are the edge states as obtained from the low-energy ${\bf k \cdot p}$ theory. (d) Behaviour of the chiral edge states Fermi velocity vanishing at the ${\bf k \cdot p}$ expansion point.
} 
\label{fig:CB}
\end{figure}

Next, we prove that the two topological classes of continuum ${\bf k \cdot p}$ Hamiltonians can be discriminated not only from their different topological invariants but also from the ensuing properties of the topologically protected edge states.  This can be immediately seen by analyzing as before the long-wavelength Hamiltonian of Eq.~\ref{eq:checkerboardkp}  on a half-plane with FBC and NBC boundary conditions and scanning for localized  wavefunctions exponentially decaying into the bulk. It can be then shown that such edge states 
 are non-dispersive and located
at the top (bottom) of the valence (conduction) band   $E= \Delta_0 \, \mbox{sign}(k_y)$ with the wave functions reading 
\begin{equation}
\Psi\propto\left [ i \, , \mbox{sign}(k_y) \right]^T (e^{\lambda{_1} \, x}\pm e^{\lambda_{2} \, x})
\end{equation}
where, similarly to the BHZ model,  the $\pm$ are for the NBC and the FBC boundary conditions respectively and the $\lambda_{1,2}$ satisfy
\begin{equation}
\lambda_{1,2}= -\frac{t\cos\phi}{2t_2} |k_y|\pm k_y \sqrt{\frac{t^2\cos^2\phi}{4t_2^2}-1}.
\label{eq:edgestatecheckerboardlambda}
\end{equation}
Fig.~\ref{fig:CB}(b) shows the energy bands in the QAH state for a checkerboard lattice ribbon.
At the the low-energy $M$ point projection, 
we find edge states precisely at the top (bottom) of the conduction band  [c.f. Fig.~\ref{fig:CB}(c)] of vanishing Fermi velocity [c.f. Fig.~\ref{fig:CB}(d)], in perfect agreement with the foregoing ${\bf k \cdot p}$ analysis . 
Most importantly, these edge states acquire a linear dispersion moving away from the low-energy ${\bf k \cdot p}$ expansion point and eventually connect the valence to the conduction band, as guaranteed by the bulk-edge correspondence. 
This feature shows the most peculiar characteristic of the {\it non-local} ${\bf k \cdot p}$ topological class: even though the topological invariant can be inferred from the low-energy bulk bandstructure, it exists a {\it non-local} bulk-edge correspondence predicting midgap edge states at different BZ points. 
This is highlighted by the fact that  right- and left-moving modes cross each other at an high-symmetry point of the one-dimensional BZ different from the low-energy BZ point.

\paragraph{Conclusions --} To sum up, we have identified two different topological classes for single-valley ${\bf k \cdot p}$ Hamiltonians of Chern insulators. In the first  {\it local} ${\bf k \cdot p}$ topological class, the momentum manifold can be one-point compactified to a Riemannian sphere, and allows for the usual  definition of the Chern invariant. 
In the second  {\it non-local} ${\bf k \cdot p}$ topological class, instead, the momentum manifold is holomorphic to a non-compact hemisphere but still allows for a one-dimensional topological invariant contained in a mapping from a circle to a circle. These two topological classes are characterized by a bulk-edge correspondence different in nature. While in the  {\it local} ${\bf k \cdot p}$ class the topologically protected edge states appears in close proximity to the low-energy points of the BZ, in the {\it non-local} ${\bf k \cdot p} $ class the bulk-edge correspondence is intrinsically non-local: midgap edge states appear away from the ${\bf k \cdot p}$ expansion point. 

\paragraph{Acknowledgements --}  CO acknowledges the financial support of the Future and Emerging Technologies (FET) programme within the Seventh Framework Programme for Research of the European Commission, under FET-Open grant number: 618083 (CNTQC).

\end{document}